\documentclass{iaus}
\usepackage{AMSgen,AMSsymb,AMSfonts,AMSbsy}
\usepackage{graphicx,natbib,natbibspacing,times}

\title[High Energy Emission from Supernova Remnants]
{High Energy Emission from Supernova Remnants}

\author[Jacco Vink]{Jacco Vink$^1$}

\affiliation{$^1$Astronomical Institute, University Utrecht,
PO Box 80000, NL-3508TA,Utrecht, The Netherlands\break
j.vink@astro.uu.nl
}

\pubyear{2005}
\volume{230}  
\pagerange{119--126}
\date{?? and in revised form ??}
\jname{"Populations of High Energy Sources in Galaxies"}
\editors{ Evert J.A. Meurs \& G. Fabbiano, eds.}
\begin{document}

\newcommand{\apss}{{Astroph. and Space Science}}
\newcommand{\apj}{{ApJ}}
\newcommand{\apjs}{{ApJS}}
\newcommand{\apjl}{{ApJ}}
\newcommand{\aj}{{AJ}}
\newcommand{\aap}{{A\&A}}
\newcommand{\aaps}{{A\&AS}}
\newcommand{\nat}{{Nat}}
\newcommand{\mnras}{{MNRAS}}
\newcommand{\prc}{{PhRvC}}
\newcommand{\araa}{{ARA\&A}}
\newcommand{\pasj}{{PASJ}}
\newcommand{\pasp}{{PASP}}
\newcommand{\npa}{{NuPhA}}
\newcommand{\iaucirc}{{IAU circ.}} 
\newcommand{\adspr}{{Adv. Space Res.}}
\newcommand{\jetp}{{JETP}}
\newcommand{\phrvl}{{PhRvL}}
\newcommand{\phrc}{{PhRvC}}
 
\newcommand{\rvmp}{{\it Rev. Mod. Physics}}

\newcommand{\xmm}{{\it XMM-New\-ton}}
\newcommand{\chandra}{{\it Chandra}}
\newcommand{\einstein}{{\it Einstein}}
\newcommand{\asca}{{\it ASCA}}
\newcommand{\rosat}{{\it ROSAT}}
\newcommand{\cangeroo}{{\it CANGEROO}}
\newcommand{\whipple}{{\it Whipple}}
\newcommand{\hegra}{{\it HEGRA}}
\newcommand{\hess}{{\it HESS}}
\newcommand{\sax}{{\it BeppoSAX}}
\newcommand{\rxte}{{\it RXTE}}
\newcommand{\osse}{{\it CGRO}-OSSE}
\newcommand{\egret}{{\it CGRO-EGRET}}
\newcommand{\integral}{{\it INTEGRAL}}
\newcommand{\glast}{{\it GLAST}}
\newcommand{\comptel}{{\it CGRO-COMPTEL}}

\newcommand{\kte}{{$kT_{\rm e}$}}
\newcommand{\net}{{$n_{\rm e}t$}}
\newcommand{\kms}{{km\,s$^{-1}$}}
\newcommand\netunit{cm$^{-3}$s}
\newcommand\pcc{cm$^{-3}$}

\newcommand\casa{{Cas\,A}}
\newcommand\msh{G292.0+1.8}
\newcommand\snts{{SN\,1006}}
\newcommand\rcwes{{RCW\,86}}
\newcommand\rxjSNR{{RX~J1713.7-3946}}

\newcommand{\tiff}{{$^{44}$Ti}}
\newcommand{\scff}{{$^{44}$Sc}}
\newcommand{\caff}{{$^{44}$Ca}}
\newcommand{\cofs}{{$^{56}$Co}}
\newcommand{\nifs}{{$^{56}$Ni}}
\newcommand{\fefs}{{$^{56}$Fe}}
\newcommand{\halpha}{{H$\alpha$}}
\newcommand{\healpha}{{He$\alpha$}}
\newcommand{\arcsec}{{$^{\prime\prime}$}}
\newcommand{\arcmin}{{$^{\prime}$}}
\newcommand{\degr}{{\symbol{23}}}
\newcommand{\msun}{{$M_{\odot}$}}
\newcommand{\ep}{{e$^+$e$^-$}}
\newcommand{\fluxunit}{{ph\,cm$^{-2}$s$^{-1}$}}

\maketitle

\begin{abstract}
This paper discusses several aspects of current research on high energy
emission from supernova remnants, covering the following main topics:
1) The recent evidence for magnetic field
amplification near supernova remnant shocks, which makes that
cosmic rays are more efficiently accelerated than previously thought.
2) The evidence that ions and electrons in some remnants
have very different temperatures, and only equilibrate through
Coulomb interactions. 3) The evidence
that the explosion that created Cas A was asymmetric, and seems to have
involved a jet/counter jet structure.
And finally, 4), I will argue that the unremarkable properties of
supernova remnants associated with magnetars candidates, suggest that
magnetars are not formed from rapidly ($P\approx 1$~ms) rotating
proto-neutron stars, but that it is more likely that they are formed
from massive progenitors stars with high magnetic fields.
\end{abstract}

\firstsection 

\section{Introduction}

Supernova remnants (SNRs) are important high energy sources.
Not only are they thought to be the major source of Galactic
cosmic rays of energies up to at least $10^{15}$~eV, they are,
until a supernova occurs in the Galaxy,
also the only Galactic sources with which can get a direct view
on supernova explosions and nucleosynthesis. 

Moreover, SNRs are sources in which interesting
physical processes take place, some of which can now
be spatially or spectroscopically resolved with the current generation
of high spatial and spectral resolution X-ray telescopes, \chandra,
and \xmm.

\begin{figure*}
\centerline{
\parbox{0.45\textwidth}{
\includegraphics[width=0.45\textwidth]{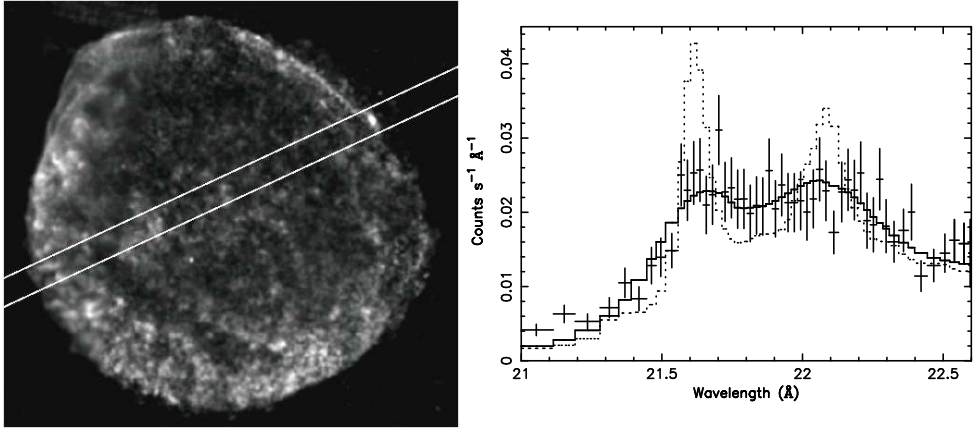}}
\parbox{0.5\textwidth}{
\includegraphics[angle=-90, width=0.5\textwidth]{sn1006_ovii_broadening.eps}}
}
\caption{On the left: Map of O\,VII emission made from
several \chandra-ACIS observations. The lines indicate
the region observed by the \xmm\ RGS instrument.
The target was the bright knot in the northeast.
Right: Detail of the RGS1 spectrum of the northeastern knot, showing
O\,VII He$\alpha$ line emission.
The dashed line is the best fit model without line broadening,
whereas the solid line shows the model including thermal line 
broadening \citep{vink03b}.
\label{fig_sn1006}}
\end{figure*}

\section{Collisionless shocks}
That the hot plasma of SNRs that we observe with X-ray telescopes
exists is somewhat of a surprise. The reason is that SNR shocks
take place in the tenuous interstellar medium, with typical densities
of $n\sim 1$~\pcc. Two body atomic collisions in such a medium are
so rare that, were it not for the generation of plasma waves, 
plasmas could move through each other without hardly any noticeable
interaction.
Because of this lack of two body  collisions (Coulomb interactions) 
SNR shocks are called collisionless.

The small collision rates in SNRs has two consequences. One of them,
non-equilibrium ionization (NEI), has been well established since the
eighties \citep[see e.g.,][for an introduction]{liedahl99}.
SNR plasma are usually in NEI, meaning that the plasma has not
yet have time to reach ionization equilibrium. In other words
the plasma is still ionizing. NEI plasmas are usually characterized
by the parameter \net, i.e. the product of electron density and
time. Plasmas with temperatures of \kte$\sim 1$~keV are out of equilibrium
when \net $\lesssim 10^{12}$~\netunit.

Less well known is that the there is considerable uncertainty
how collisionless plasmas are heated by shocks, and in particular
whether this leads immediately downstream of the shock
to equal temperatures for all particles (electrons, ions) involved.
Basic considerations, such as conservation of mass, momentum and energy,
lead to the Rankine-Hugoniot relations, but it is not clear whether 
collisionless shock heating leads to equal temperatures for
all particles, or whether it heats different plasma elements
to different temperatures.
In the latter case one expects that the temperature of each species
is proportional to $kT_i \propto m_i v_s^2$, with $m_i$\ the particle
mass and $v_s$\ the shock speed. Far downstream of the shock the
particles will eventually equilibrate their energies. Also the Coulomb
equilibration of electrons and ions is governed by the parameter \net.
Equilibration is reached when \net $\lesssim 10^{12}$~\netunit.

Over the last decade several observations in
the optical \citep{ghavamian01,ghavamian03}, 
UV \citep{raymond95,laming96,korreck04}, 
and X-rays \citep{vink03b} have been used to address this issue.
They all rely on the fact that ion temperatures can be measured
by observing thermal Doppler broadenings. For SNRs this can only
be done for the limb brightened edges of the shells, because far inside
the shell, bulk motions in the line of sight dominate the line broadening.

In Fig.~\ref{fig_sn1006} X-ray evidence is shown that for SN 1006
the ion temperature is much hotter, $kT_{\rm O VII} \sim 500$~keV, than
the electron temperature \kte$ \approx 1.5$~keV.
Of course what is normally measured is the electron temperature,
as this determines the X-ray continuum shape and emission line ratios.

SN1006 is a very suitable target for such a study, as its plasma
is very far out of equilibrium, $\log$\,\net$ = 9.5$.
However, optical, \halpha, measurements of line broadening do not rely
on low \net\ values, as \halpha\ emission only occurs very close
to the shock front. Using \halpha\ measurements of several
SNRs, \citet{rakowski03}
have shown that slow equilibration of temperatures is probably a
function of shock velocity or Mach number, as only the fastest shocks
appear to have substantial differences between electron and proton 
temperatures.

\begin{figure*}
\centerline{
\parbox{0.5\textwidth}{\vbox{
\includegraphics[width=0.5\textwidth]{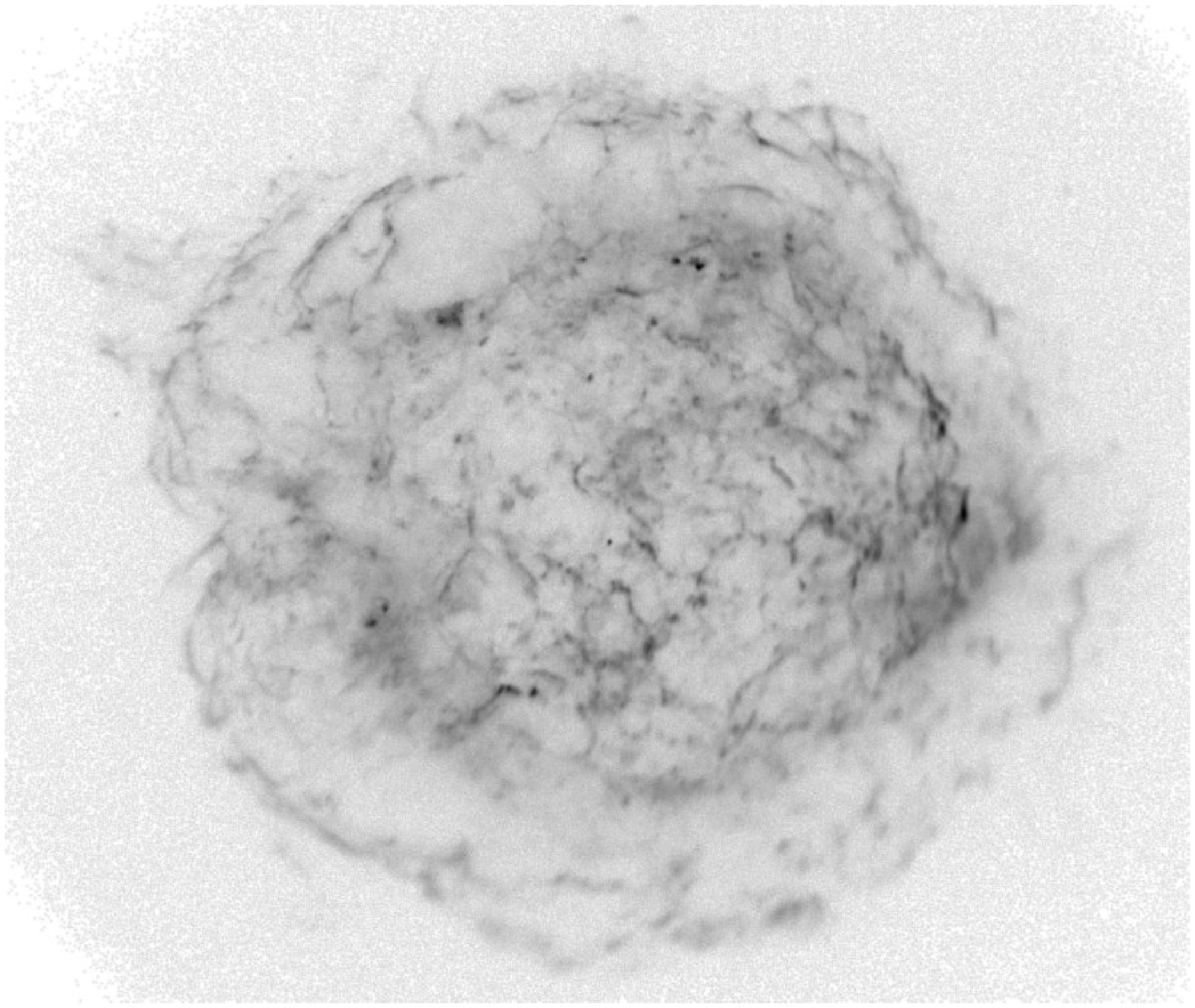}
    \vskip 5mm}
}
\parbox{0.5\textwidth}{\vbox{
     \includegraphics[width=0.5\textwidth]{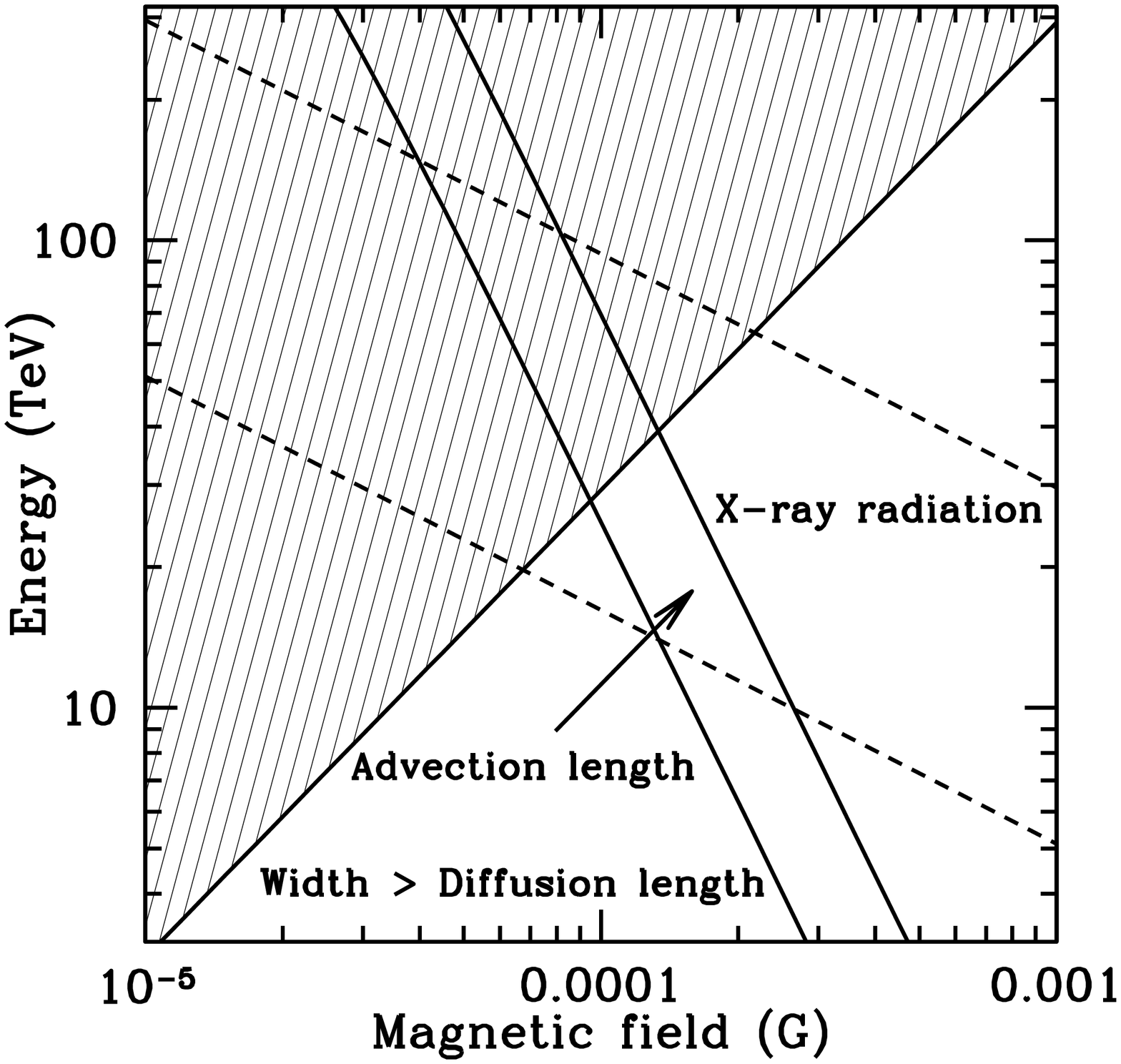}
     \vskip 0mm
  }
}
}
  \caption{A deep \chandra\ image of Cas A \citep{hwang04}
in the 4-6 keV continuum band (left).
Note the thin filaments, marking the border of the remnant
(NB the point spread function is not
uniform). The remnant has a radius of about 2.5\arcmin.
Right: Determination of the maximum electron
energy versus magnetic field strength
for the region just downstream of \casa's shock front, as determined
from the thickness of the filaments. The shaded area
is excluded, because the filament width cannot be smaller than the minumum
possible diffusion length \citep[c.f.][]{vink03a}.
\label{bfield}}
\end{figure*}

\section{Cosmic ray acceleration and magnetic field amplification}
One of the important findings of \chandra\ concerning SNRs was that
with its CCD detectors it was able to pick out thin, non-thermal
X-ray emitting filaments \citep{hwang02,gotthelf01a}.
\citet{vink03a} showed that these filaments probably emit
synchrotron emission, with electron energies $\gtrsim 10$~TeV.
The narrow widths of these filaments are then best interpreted as the
result of synchrotron losses.

The reason is that the plasma downstream of the shock sweeps the
relativistic particles away from the shock. At the same time
the electrons rapidly lose energy, so that at some point away
from the shock front the electrons only emit synchrotron
radiation at energies below the X-ray band.
For standard shock high Mach number shocks
the plasma velocity with respect to  the shock front is given by 
$\Delta v = \frac{1}{4} v_s$. So for the width of the filaments
one can write $Delta r = \frac{1}{4} v_s \tau_{loss}$,
with the synchrotron loss time given by $\tau_{loss} = 635/(B^2E)$.
In order to disentangle the electron energy $E$, and the average
downstream magnetic field strength $B$\ one has to use
the fact that the peak photon energy as a result of synchrotron
radiation is $\epsilon = 7.4 E^2B$~keV.
Fig.~\ref{bfield} shows graphically what for Cas~A the possible
values for $B$ and $E$ are. It turns out that the magnetic
field is high $B = 200-500$~$\mu$G for \casa, but also for other young SNRs 
\citep[e.g.][]{bamba05,ballet05},
is much stronger than might be expected if the magnetic is just
the shock compressed mean Galactic field.

This may be surprising, but it is a nice confirmation of recent theoretical
work that indicates that strong cosmic ray streaming close to
fast SNR shocks may lead to non-linear amplification of magnetic fields
\citep{bell01,bell04}.

In fact this solves a piece of the puzzle concerning cosmic ray acceleration.
SNRs were for a long time thought to be the most plausible sources
of Cosmic Rays up to or beyond $3\times10^{15}$~eV, at which energy
the cosmic ray spectrum has a break. However, with only mean
Galactic magnetic field values SNRs are not able to efficiently accelerate
particles up to even $10^{14}$~eV \citep{lagage83}.
It looks now that \chandra\ has solved this problem.

Although, X-ray observations show that one necessary ingredient,
large, turbulent magnetic fields are present \citep[see also][]{vink04dxx},
there is still no direct evidence that SNRs accelerate also ions
up to high energies. Note that ions are the main ingredient
of the cosmic rays that bombard the earth atmosphere.
In that respect the many SNRs that have been observed with Cherenkov
telescopes, in particular H.E.S.S. \citep[][]{aharonian04}, are very promising.
The reason is that collisions of relativistic ions result in the
production of pions, and $\pi^0$\ particles decay into two photons, 
giving rise to $\gamma$-ray emission.
However, also inverse Compton scattering of background photons by relativistic
electrons produces $\gamma$-ray emission. In that case X-ray
and Cherenkov $\gamma$-ray telescopes may observe the same
electron cosmic ray population. It has not yet been resolved which
mechanism is responsible for the $\gamma$-ray emission from SNRs.

\begin{figure}
\centerline{
\includegraphics[width=0.45\columnwidth]{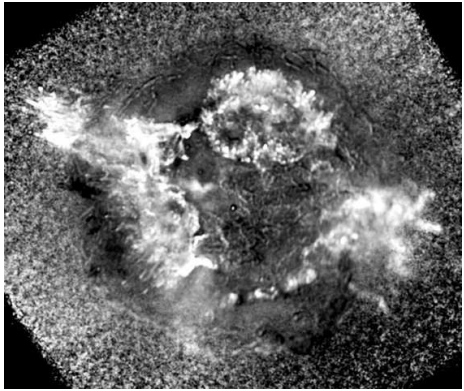}
}
\caption{
Image based on the deep \chandra\ observation of \casa,
which has been processed to bring out the jet/counter jet structure
\citep{vink04a,hwang04}.
(Credit: NASA/CXC/GSFC/U.Hwang et al.)\label{jet}}
\end{figure}

\section{Asymmetric supernova explosions: a link with GRBs?}
The origin of cosmic rays may be an almost century old problem,
but an equally fascinating, but more recent problem is the
nature of gamma-ray bursts (GRBs).
It is becoming more and more clear  that long duration GRBs are probably
associated with core collapse supernovae of subclass Type Ibc
\citep{stanek03}. However, the mechanism that generates the
powerful relativistic jets that we observe as GRBs is not well
known. The collapsar model \citep{macfadyen01}
is one of the most popular models. In this model the stellar core
collapses into a black hole that accretes matter, and generates
jets. An alternative model is magneto-rotational
jet formation \citep{akayima03}. 

In this light it is interesting that it was recently discovered
that 
the bright Galactic SNR Cas A seems to have exploded with a jet/counter jet.
A normal X-ray image of Cas A does not immediately reveal this,
but  dividing a narrow band image dominated by Si XIII line
emission by a narrow band image with Mg XI line emission
brings out a clear jet/counter jet structure 
\citep[Fig.~\ref{jet}][]{vink04a,hwang04}.
The spectra of the jet reveal
an apparent absence of Ne and Mg. The dominant elements seem to be 
Si, S, and Ar, but some Fe seems also present.
The emission measure of the jet combined with the average velocity of
the plasma suggest quite a high kinetic energy in the jet, 
$\sim 5\times10^{50}$~erg, about 25\% of the total explosion energy.

So in terms of the total explosion energy the jets seem to contain
as substantial, but not a dominant fraction of the energy.
The jets seem not have been relativistic jets, like those of GRBs. 
Nevertheless,
perhaps the same underlying mechanism produces both types of jets.
In that respect it is interesting that Cas A is likely the result
of a Type Ib explosion, i.e. it belonged to the same supernova subclass
with which GRBs are associated.
However, it is unlikely that the collapsar model in its present form
is responsible for the jets in Cas A, because there is the simple
fact that the explosion appears to have resulted in the formation
of a recently detected neutron star \citep{tananbaum99} 
rather than a black hole.

Interestingly, not only the presence of a jet/counter jet makes
Cas A an interesting SNR from the point of view of the explosion mechanism.
Equally interesting is that Fe-rich knots seem to have been ejected with
greater speed than the Si-rich material synthesized further away from
the core. This is clear from the presence of Fe-rich knots in the southeast
of the remnant,  outside of the main Si-rich shell \citep{hughes00a,hwang03}.

In the north the Fe is projected to the  inside of the Si-rich shell. 
However, this appears to be a projection effect, because the measured 
Doppler velocities of Fe in the north is higher than Si \citep{willingale02}.
It is not clear how much of the Fe in Cas A is
still unshocked, but some of the shocked Fe must have been ejected with
velocities of up to 7800~\kms.
There is no obvious symmetry to the Fe-rich ejecta, so their emergence
is probably related to hydrodynamical instabilities close to the core
of the explosion \citep{kifonidis03}.

The 3D morphology of Cas A as reconstructed from the Doppler imaging
obtained from \xmm\ gives further evidence that the explosion
was from spherical 1\citep{willingale02}. 
Apart from the jets and Fe-rich knots, the ejecta
can best be described by a donut shape. This morphology also
appears to describe best the high resolution spectroscopy data 
of the Small Magellanic Cloud remnant 1E0102.2-7219, obtained
by the grating spectrometers of \chandra\ \citep{flanagan04}.

\begin{figure*}
\centerline{
\parbox{0.45\textwidth}{
\includegraphics[width=0.45\textwidth]{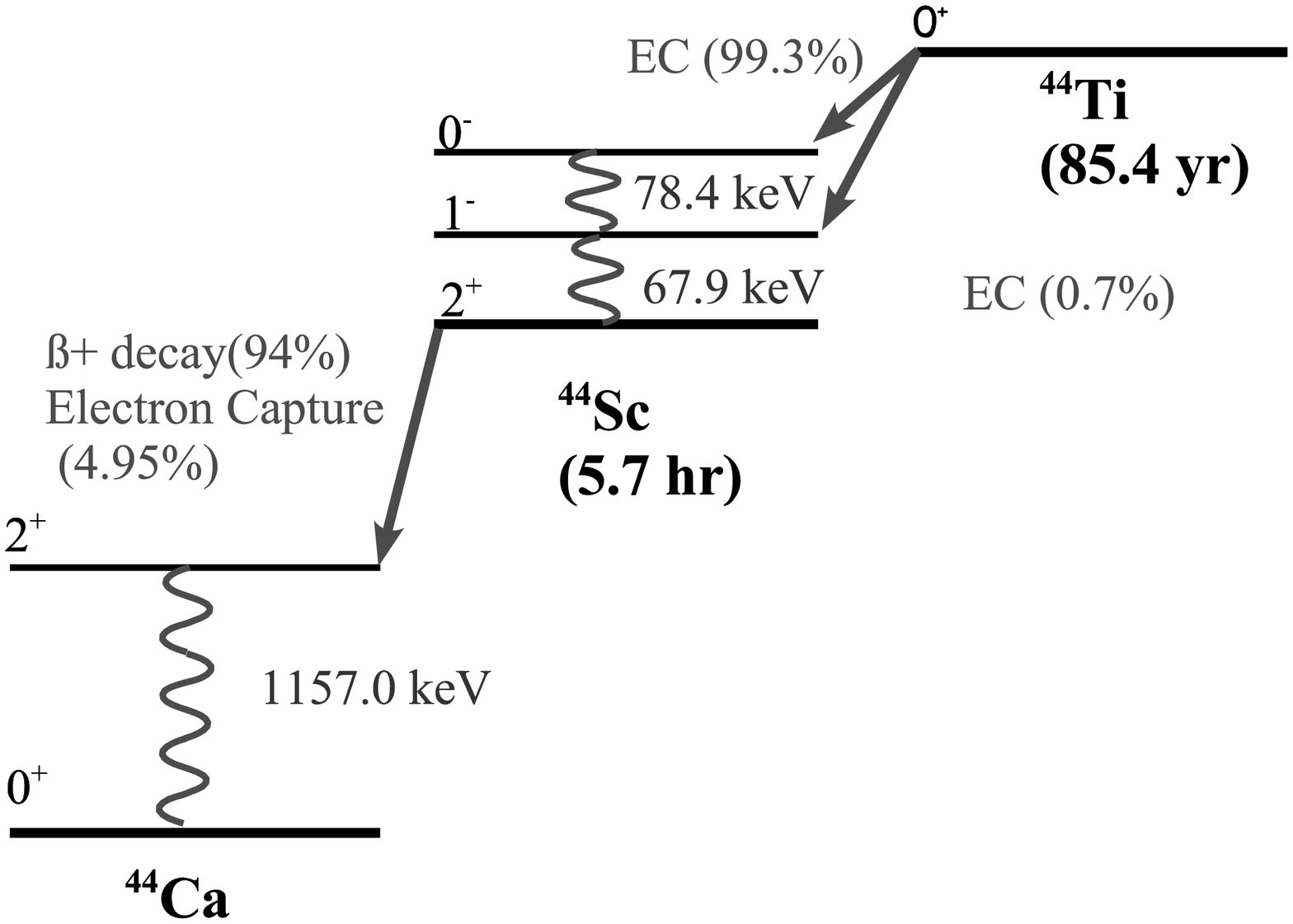}}
\parbox{0.45\textwidth}{
\includegraphics[angle=-90,width=0.45\textwidth]{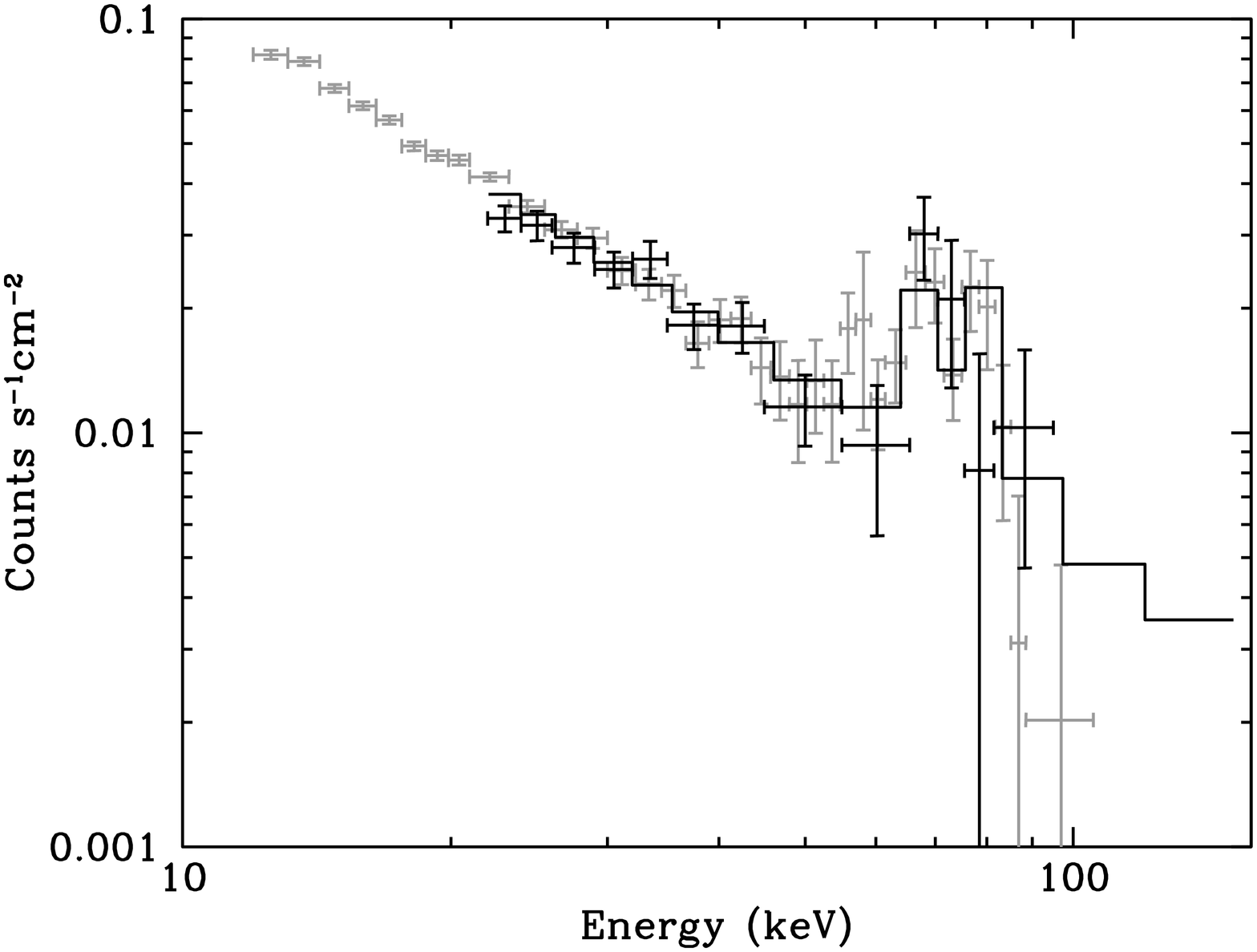}}
}
\caption{
\label{fig_ti44}
Left: The radio active decay scheme of \tiff. Right: hard X-ray spectrum,
as observed by \sax\ (in gray) and \integral, both showing clear
signs of emission around 68~keV and 78~keV due to \tiff\ decay lines
\citep{vink05a}.
}
\end{figure*}

There is one other observational fact concerning Cas A that
may point to an intrinsically asymmetric explosion: the
presence of \tiff, which suggest a high initial
\tiff\ yield of $\sim 10^{-4}$~\msun.
\tiff\ is exclusively an explosive nucleosynthesis product,
and is synthesized close to the core of the explosion.
It is an alpha-rich freeze out product \citep{arnett96},
and as such very sensitive to explosion energy and explosion
asymmetries. 

The decay of \tiff\ ($\tau=86$~yr)
is accompanied by three strong $\gamma$-ray lines
(Fig.~\ref{fig_ti44}), which have been detected by \comptel\ 
\citep[the 1157 keV line,][]{iyudin94}, and \sax\ 
\citep[the 68 and 78~keV lines,][]{vink01a}.
The fact that the line emission is caused by radio-active decay
makes that with \tiff\ we can also probe the unshocked \tiff\ ejecta.
\casa\ is therefore a target of ESA's $\gamma$-ray observatory INTEGRAL.
Preliminary results of the observations have been published in
\citet{vink04a} (Fig.~\ref{fig_ti44}).

\section{The remarkable magnetars and the unremarkable
supernova remnants in which they reside}
Core collapse supernovae obtain their energy from the gravitational
potential energy released when the core of a massive star collapses.
The classical view is that core collapse would either
result in the creation of a radio pulsar with typical
magnetic fields of $B_{dip} \sim 10^{12}$~G, or otherwise
in the formation of a black hole.
However, over the last decade it has become clear
that some neutron stars appear to have very high magnetic fields,
up to $B_{dip}\sim 10^{15}$~G. Such neutron stars
are called ``magnetars''. It is thought that two classes of X-ray
pulsars are manifestations of magnetars.
Depending on the presence or absence of (soft) $\gamma$-ray flashed,
these pulsars are either labeled Soft-gamma-ray-repeaters (SGRs),
or Anomalous X-ray Pulsars (AXPs) \citep[see][for reviews]{kaspi04,woods04}.

It is not clear what causes the creation of a magnetar, but two mechanisms
have been proposed. Perhaps the most popular explanation is that magnetars
are created during the core collapse of massive stars with a high angular
momentum. This results in the formation of a proto-neutron star
which rotates non-uniformly with an average rotation period
close to the break up limit $P \sim 1$~ms. This allows for
the efficient operation of an $\alpha-\Omega$ dynamo, which rapidly
amplifies the magnetic field \citep{duncan92}.
Once the high magnetic field is in place, a magnetar will lose  most of
its angular momentum due to magnetic breaking in less than a few hunderd
seconds, i.e. during the supernova explosion self \citep[e.g.][]{thompson04}.
This means that  most of the rotational energy will be rapidly pumped into
the supernova ejecta. A neutron star with $P=1$~ms has a rotational
energy of $10^{52}$~erg, which should be compared to an average
core collapse supernova energy of  $10^{52}$~erg.

\begin{figure*}
\centerline{
  \parbox{0.45\textwidth}{
    \includegraphics[width=0.45\textwidth]{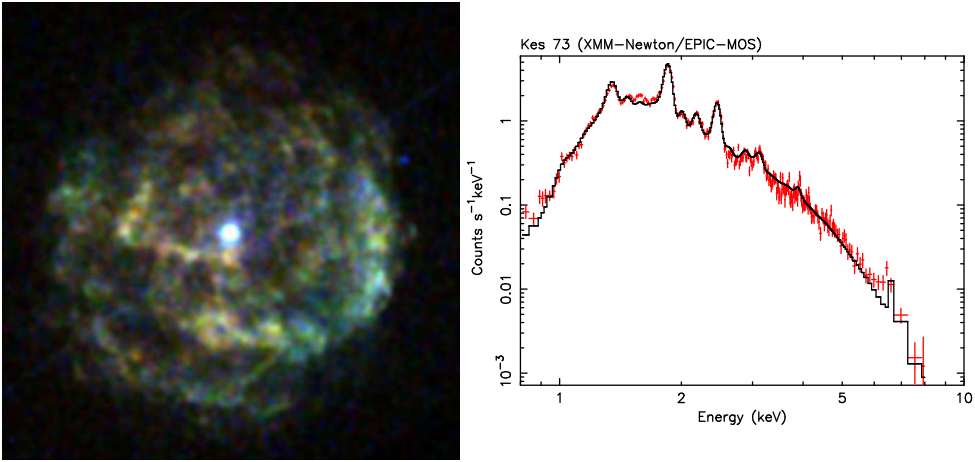}
  }
  \parbox{0.5\textwidth}{
    \includegraphics[angle=-90,width=0.5\textwidth]{kes73_mos12_sedov.ps}
  }
}
\caption{
\label{fig_kes73}
Left: \chandra\ image of Kes 73. The AXP itself is 
highly saturated and shows up as a blueish spot in the center of the
remnant. 
Right: \xmm\ EPIC-MOS spectrum
of Kes 73, with the best fit Sedov model shown as a solid line.
From this Sedov model \citep{borkowski01b}
one can estimate the total kinetic energy
of the supernova (Vink et al. in preparation).
}
\end{figure*}

What interests us here is the fact that of the dozen or so magnetar candidates,
three are associated with bright SNRs: Kes 73, N49, and CTB 109.
So one would expect that the kinetic energy of these SNRs is exceptionally
high, as the ejecta have been energized by the rapid spin down of the
rapidly rotating magnetar.
However, the kinetic energies of these SNRs are unremarkable:
A study of the energetics with \xmm-EPIC spectroscopy reveals that
the kinetic energies of  Kes 73, N49, and CTB 109 are resp.
$0.8\times 10^{51}$~erg, $2.0\times 10^{51}$~erg, and $0.7\times10^{51}$~erg 
\citep[Vink \& Kuiper, 2005, MNRAS submitted, and][]{sasaki04}. 
This is far short
of the $\sim3\times10^{52}$~erg expected if magnetars owe their existence
to the operation of an $\alpha-\Omega$ dynamo \citep{duncan92}.
In fact one can put a lower limit on the initial spin period, by
equating the observed SNR energies to the initial rotational energy
of the pulsar. This gives $P_i > 5.6 \sqrt(E_{SNR}/10^{51}~{\rm erg})$~ms,
which is closer to the classical initial spin period of radio pulsars, 10~ms,
than to the break-up limit of a neutron star.

So the unremarkable energies of SNRs associated with magnetars imply that
magnetars are the result of collapses of the cores of massive
stars with high magnetic fields \citep[see also][]{ferrario05}, 
rather than from stars with a high angular momentum.

\section{Summary}
I have shown that X-ray studies of SNR provides us with important
information on collisionless shock physics, cosmic ray acceleration
and magnetic field amplification by SNRs.

Moreover, by studying SNRs we can learn about the details of the
supernova explosions that caused them. For example,
the kinematics and spatial distribution of metals in Cas A reveal 
that the explosion was intrinsically asymmetric, and was accompanied
by the emergence of a jet/counter jet system. 
And I have discussed that the most remarkable property of SNRs associated 
with magnetars is that they are unremarkable:
Their energies are similar to those of other SNRs, 
which suggests that magnetars were {\em not} formed
from proto-neutron stars with period close to 1~ms.



\end{document}